\documentclass[preprint]{aastex}
\usepackage{psfig,emulateapj5}

\shorttitle{Local Counterparts to Compact Blue Galaxies}
\shortauthors{Barton \& van Zee}

\begin{document}
\slugcomment{To appear in ApJ Letters}
\title{Possible Local Spiral Counterparts to Compact Blue
Galaxies at Intermediate Redshift}
\author{Elizabeth J. Barton and Liese van Zee}
\affil{Herzberg Institute of Astrophysics, National Research Council of
Canada, 5071 W. Saanich Rd., Victoria, BC, Canada V9E 2E7 
(email: Betsy.Barton@hia.nrc.ca)}

\begin{abstract}

We identify nearby disk galaxies with optical structural parameters 
similar to those of intermediate-redshift compact blue galaxies.
By comparing \ion{H}{1} and optical emission-line widths, we
show that the optical widths substantially
underestimate the true kinematic widths of the local galaxies.
By analogy, optical emission-line widths
may underrepresent the masses of intermediate-$z$ compact objects.
For the nearby galaxies, the compact blue morphology is the result of
tidally-triggered central star formation:
we argue that interactions and minor mergers may
cause apparently compact morphology at higher redshift.

\end{abstract}


\section{Introduction}

Blue galaxies are abundant at intermediate redshift.
Many are luminous, with distinct 
irregular or compact morphology \citep{P97} 
and high rates of star formation.  
Number counts indicate they are the most strongly evolving 
population at intermediate redshift [\citet{CFRS}; \citet{E97} and 
references therein].  Thus, a complete understanding of galaxy evolution
from $z=1$ to the present requires understanding what
stage of evolution the luminous blue galaxies represent.

The most extreme luminous blue galaxies at intermediate 
redshift are the compact narrow emission-line galaxies 
\citep[CNELGs hereafter;][]{K94,K95,G96,G97,G98}.
With small half-light radii (1~--~3.5 kpc) and narrow
emission-line velocity widths 
($35 \lesssim \sigma < 126$~km~s$^{-1}$ from spatially unresolved 
spectra),  CNELGs resemble few nearby objects.
Their nature is currently heavily debated.
Several authors \citep[e.g.,][]{K94,K95} 
argue that the narrow emission-line widths 
of CNELGs indicate small masses; \citet{K95} suggest CNELGs
are the starbursting spheroidals faded by 4~--~7 
magnitudes at the current epoch.  However, spectroscopic
studies of field galaxies at intermediate redshift \citep{CL00}
and of compact objects in particular \citep{KZ99,H00} find high
metallicities for some blue galaxies, suggesting that
some, including some CNELGs, may be bulges forming with
central bursts of star formation in luminous spirals.

At the heart of the debate is the question of
the masses of the luminous compact blue galaxies.
Additional kinematic information for the CNELGs, 
such as \ion{H}{1} linewidths or 
stellar kinematics, would dramatically improve our knowledge of these
systems.  However, 21cm measurements are nearly impossible at
intermediate $z$ and any older stellar populations of 
CNELGs are too faint to measure stellar kinematics.  One remaining 
avenue is study of local counterparts with properties similar to
those of the CNELGs \citep[e.g.,][]{P01}.

Recently, \citet{B01} identified a class of possible nearby 
counterparts to CNELGs and other luminous compact galaxies. 
The local galaxies are 
spirals in pairs with blue star-forming centers, 
small half-light radii, and anomalously narrow emission-line 
rotation widths.  In these cases, concentrated line emission probably 
results from gas infall after a close galaxy-galaxy interaction, a process 
similar to central star formation triggered in minor mergers
\citep{MH94, MH96}.
Here, we argue that observations of these possible counterparts strongly
support the hypothesis that at least some luminous blue compact galaxies
at high redshift are disk galaxies with central star formation.
In Sec.~2, we detail the evidence
that these local objects are counterparts
to compact blue galaxies at intermediate redshift. In Sec.~3,
we present \ion{H}{1} synthesis observations of two of the
four galaxies which show that even resolved major axis emission-line 
rotation curves underrepresent the full kinematic widths of the
local galaxies by more than a factor of two, allowing 
the possibility that in spite of their
narrow emission-line widths, CNELGs are the centers of
intrinsically massive systems.  We conclude in Sec.~4.  
We use H$_0 = 50$~km~s$^{-1}$~Mpc$^{-1}$ and, where applicable, 
q$_0 = 0.1$.

\section{Local Disk Galaxies with Blue Centers}

If luminous blue compact morphology is a brief transitional stage
in massive galaxies,
local counterparts may exist among
galaxies that are rapidly
evolving at the current epoch, such as galaxies in pairs.
Many studies of the stellar populations of paired galaxies
identify substantial new star formation triggered by 
interactions \citep[e.g.,][]{LT78,K87}.  
Recently, \citet{BGK00a} studied a large sample 
of galaxies in pairs selected from the CfA2 redshift survey
based only on separation on the sky ($\leq 50$~h$^{-1}$kpc) and
in redshift ($\leq 1000$~km~s$^{-1}$).  
New spectra of 502 of the galaxies show that
indicators of recent star formation, such as H$\alpha$
equivalent width, correlate with pair separation
on the sky and in redshift, providing unambiguous evidence
that close galaxy-galaxy passes trigger central star formation 
\citep{BGK00a}.
Subsequent $B$ and $R$ 
photometry of 190 galaxies in the sample 
indicates that

\psfig{figure=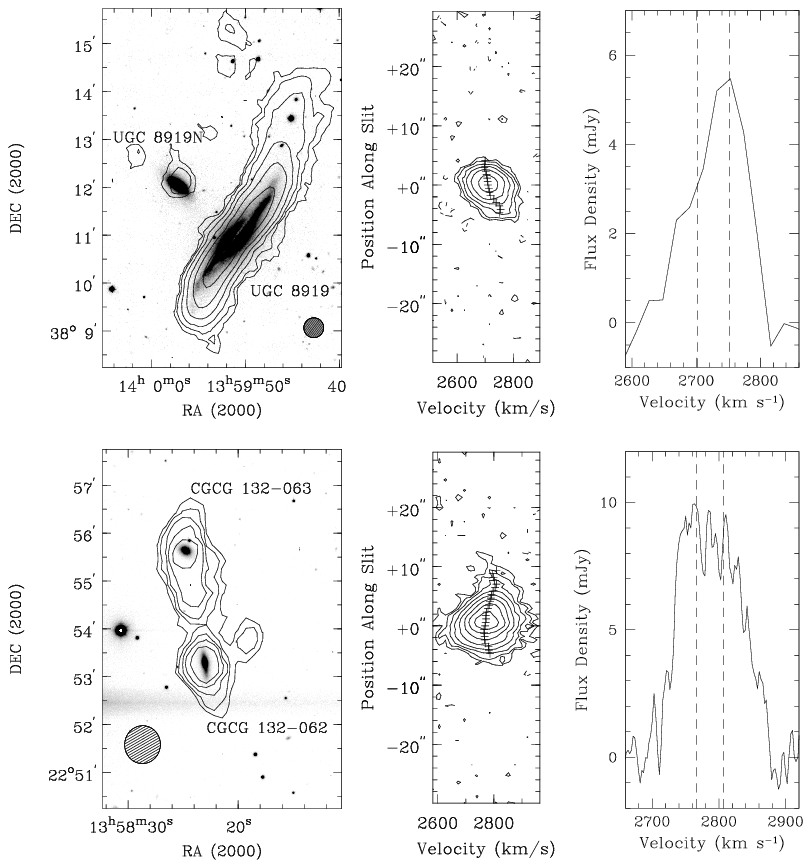,bbllx=180pt,bblly=510pt,bburx=620pt,bbury=770 pt,clip=t}
\figcaption{Images, rotation curves, and \ion{H}{1} profiles
of two candidates for counterparts to CNELGs selected from galaxies in pairs.  
The targets are UCG 8919N and CGCG 132-062.  The left panels show B
images (greyscale) and \ion{H}{1} distributions (contours).  The middle panels
show the H$\alpha$ emission along the major axes of the targets, on an
expanded spatial scale; the emission
is centrally concentrated.
Crosses mark the rotation curve derived from cross correlation analysis.  The
right panels show ``single-dish'' \ion{H}{1}  profiles
for the local counterparts (solid lines); 
vertical dashed lines denote the optical rotation widths.  
\vskip 10pt
\label{fig:fig1}}

\noindent  
flux from the new stars can dominate even
the $R$-band light in the centers of the galaxies \citep{BGK00b}.
This strong centralized star formation renders many of 
the paired galaxies good candidates for counterparts to the 
intermediate-$z$ systems.  Because the CNELGs
and the other luminous compact objects have narrow emission lines,
the emission-line kinematics provide the true test. 

\citet{B01} study the resolved emission-line rotation curves of
90 galaxies in the pair sample.  
Most lie on the Tully-Fisher \citep[T-F hereafter]{TF77} relation 
for non-interacting galaxies \citep[e.g.,][]{C97}.
However, they identify 
seven galaxies which,
like the compact objects observed at higher redshifts, 
appear significantly  ``overluminous''  
for their measured emission-line widths.  

\subsection{Optical Observations of the Local Tully-Fisher Outliers}

Four of the ``overluminous'' T-F outliers
have properties similar to those of the compact blue objects selected 
from the Hubble deep field \citep{P97}.  
[The anomalous properties of the remaining three outliers 
probably result from tidal and gasdynamical distortion \citep{B01}.]
Table~\ref{tab:table} lists the optical 
properties of the four local galaxies measured from $B$ and $R$
images and moderate-resolution major axis spectra;
\citet{B01} describe the data in more detail.
Three of the four are clearly disk galaxies; NGC~2719A is 
somewhat amorphous.
Fig.~\ref{fig:fig1} shows the images and rotation curves for  
UGC~8919N and CGCG~132-062, the two galaxies with
\ion{H}{1} synthesis observations.
We measure total magnitudes and correct them for internal
extinction as described in \citet{B01}.
For direct comparison with the higher redshift objects, 
we use circular apertures to measure the $B$-band 
half-light radius, R$_{\rm e}$.  

Table~\ref{tab:table} includes two different measures of
the velocity widths of the galaxies from optical spectroscopy,
V$_{2.2}$ and $\sigma$.
V$_{2.2}$ is the spatially resolved 
emission-line rotation curve velocity width from
\citet{C97}. It is the full width of
the rotation curve at 2.15 disk scale lengths, 
from a model fit to the data;  V$_{2.2}^{\rm c}$ is corrected
for observational effects, such as inclination, as in \citet{B01}.
The small measured velocity widths
reflect the fact that the
strongest line emission is not spatially extended (see Fig~1).

The second velocity width measure, $\sigma$, is more directly comparable
to existing measurements for higher-redshift objects.
Because it is difficult to resolve the intermediate-z galaxies
from the ground, the existing intermediate-z emission-line
measurements of the compact objects include flux from the whole galaxy.
In addition, the quoted linewidths are usually velocity dispersion
measurements of Gaussian fits to
the [OII], [OIII], or H$\beta$ emission lines \citep{K95,G96,G97,P97}.
To approximate these measurements, 
we collapse the H$\alpha$ flux along the entire major axis and fit
a Gaussian function to the H$\alpha$ line to measure $\sigma$.
We measure the resolution of each observation
from Gaussian fits to multiple sky lines in
the extracted region.  The resolutions range from 1.4~--~1.5~\AA\ 
FWHM ($\sigma_{\rm RES} = 26 - 29$~km~s$^{-1}$ at H$\alpha$).  
We apply the prescription of \citet{G97} to account for 
instrumental resolution.

\subsection{Comparison with Luminous Blue Compact Galaxies at 
Intermediate Redshift}

The intrinsic masses of intermediate-redshift compact 
objects are uncertain.  Bulk
scaling parameters are frequently the only available measures.
Fig.~\ref{fig:re_sigma} shows an example of the use of these
parameters; following \citet{G96}, 
we plot the half-light radii and velocity widths
of compact objects 
\citep[solid points: ][and references therein]{K94,K95,G96,G97,G98,P97}.
The open points are the local \citet{FreiSample} 
sample, which consists largely of massive galaxies; half-light
radii are from \citet{Ber00} and $\sigma$ comes
from approximate values of $\sigma$ based on \ion{H}{1} measurements 
($\sigma_{\rm RADIO}=W_{50}/2.35$) compiled from the
RC3 \citep{RC3} and  NED.\footnote{The 
NASA/IPAC Extragalactic Database (NED) is operated by the Jet Propulsion 
Laboratory, California Institute of Technology,
under contract with the National Aeronautics and Space Administration. }  
The four asterisks are the targets of this study, with 
$\sigma$ measured from the optical emission-line
widths (see $\sigma$ in Table~\ref{tab:table}).
This analysis, and similar plots of M$_{\rm B}$ vs. $\sigma$ and M$_{\rm B}$
vs. R$_{\rm e}$, place the local targets of the present study in
the same range as many of the compact Hubble deep field objects
\citep{P97}; they are less luminous than typical
CNELGs.

The compact objects and CNELGs fall to the lower left
of the majority of the Frei sample in Fig.~\ref{fig:re_sigma}. 
Although  

\psfig{figure=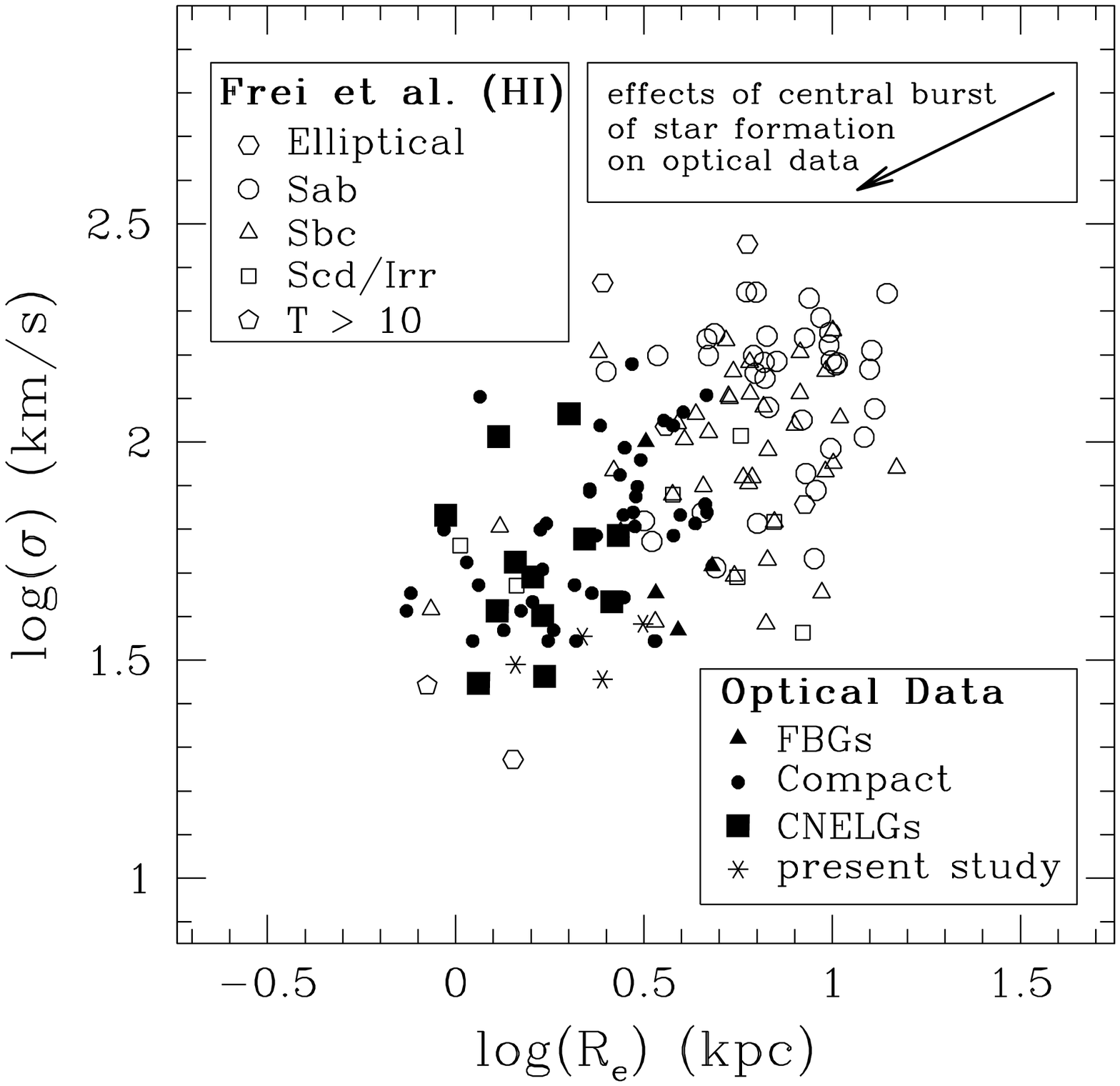,height=3.4in}
\figcaption{Structural parameters of intermediate-$z$
compact objects (filled points), local galaxies (open points),
and the proposed local counterparts to compact blue
galaxies (asterisks).
The overall trend is that smaller galaxies have smaller velocity
dispersions.  However, the arrow shows the effects of
a central burst of star formation, which can 
``artificially'' lower {\it both} the half-light radius
and the optical emission-line width of the galaxy.
\vskip 10pt
\label{fig:re_sigma} }

\noindent
these scaling parameters
seem to indicate that they are dwarf-like objects,
this argument may be entirely flawed.
The structural parameters (e.g., R$_{\rm e}$)
and the optical emission-line kinematics (both $\sigma$ and
V$_{2.2}$) are heavily influenced by regions of 
current and recent star formation; Fig.~\ref{fig:model} 
shows a simple model for the effects of 
a bulge-forming central starburst.
We model the disk with an exponentially decreasing star 
formation rate ($\tau = 4$~Gyr).  
At 7 Gyr (i.e., at intermediate redshift), 
a starburst forms an exponential bulge. 
We use the Bruzual \& Charlot (1996, in preparation) star-formation
models normalized for a bulge-to-disk ratio of 0.1 at the
present day (14 Gyr here).
The top panel shows the $B$-band half-light radius of the galaxy,
adopting a bulge half-light radius R$_{\rm e,b} = 0.5$~kpc, typical
for exponential bulges (Carollo 1999), 
and a disk half-light radius of 12.5~R$_{\rm e,b}$,
typical for late-type spirals (Courteau, de Jong, \& Broeils 1996).
The forming bulge dominates the flux and dramatically reduces
the half-light radius.  
During bulge formation the system looks like the luminous
compact, blue galaxies at intermediate redshift.  
With the additional assumptions 
of a fixed maximal-disk rotation curve, an inclination of 60$^{\circ}$, and
a linear dependence between star formation rate and H$\alpha$ flux, the
bottom panel shows the decrease in the measured linewidth during the burst
\citep[see also][]{LH96,BM01}.
In summary, Fig.~\ref{fig:model} shows that central star 
formation in the amount required to form a small bulge in a 
single burst can lower the measured half-light radius of a 
galaxy by a large amount (0.86 dex in Fig.~\ref{fig:re_sigma}) 
and can simultaneously lower the measured $\sigma$, by $> 0.2$ dex,
shifting the object in Fig.~\ref{fig:re_sigma} as shown by the arrow.

\psfig{figure=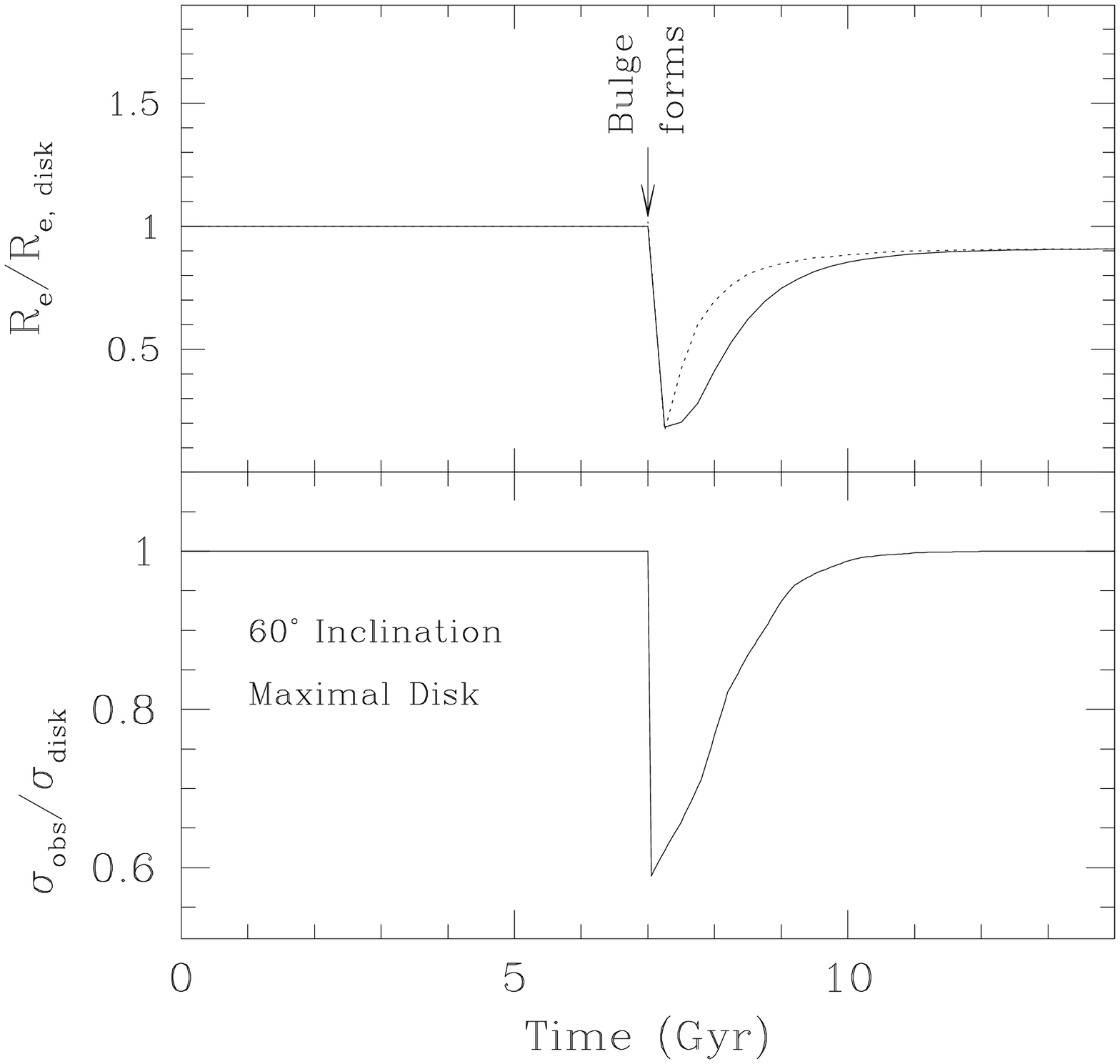,height=3.4in}
\figcaption{
A simple model for how a bulge-forming
central burst of star formation, either instantaneous (dotted line)
or extended ($\tau = 500$ Myr, solid line), could temporarily
decrease both the half-light radius ({\it top})
and $\sigma$ ({\it bottom}) of a spiral galaxy.  The model
follows the disk from 14~Gyr ago through a bulge formation event at 7~Gyr
to the present day.  See the text for a description.
\vskip 10pt
\label{fig:model}}

\section{\ion{H}{1} Observations: A Better Probe of the Kinematics}

Our local counterparts allow us to test the effects of 
central star formation on $\sigma$ by
comparing emission-line velocity widths in compact objects
with \ion{H}{1} observations \citep[see also][]{KG00, P01}.
Spatial separation of the galaxies in the pair required
\ion{H}{1} synthesis observations.  We present 
Very Large Array\footnote{The Very Large Array is a facility 
of the National Radio Astronomy Observatory.
The National Radio Astronomy Observatory is a facility of the 
National Science Foundation, operated under a cooperative 
agreement by Associated Universities Inc.}
observations of two of the four objects in this study.
We reduced archival observations of UGC 8919N originally obtained
by J. Chengalur as part of a set of observations of UGC 8919, the larger 
galaxy in the pair. The archival data include both C and D-configuration 
observations obtained on 1993 July 23 and 
1995 March 9.  They totaled 257 minutes on--source in the C configuration 
and 118.5 minutes on--source in the D configuration, 
in 4IF mode with a bandwidth of 3.125 MHz. 
The two pairs of IFs were offset in frequency to provide full velocity 
coverage of UGC 8919.  
We observed CGCG 132-062 with the VLA
in D configuration on 2000 August 18. The
observations were conducted in 2AD mode with a frequency coverage
of 1.56 MHz and on--line Hanning smoothing.  We spent a total of 191 minutes
on--source, with interspersed observations of phase and flux calibrators.

We followed standard data reduction practices, using tasks in
the {\small \rm AIPS} package to flux and phase calibrate
the {\it u--v} data \citep[see, e.g.,][for a full description of
our standard reduction methods]{vSS98}.
{\small CLEAN} data cubes were created with the {\small \rm AIPS} task 
{\small \rm IMAGR} using robustness parameters from 5 
(natural weight) to 0.5.  
Using the {\small \rm GIPSY} software package, we constructed moment maps
from clipped cubes consisting of regions with signal 
greater than 2 times the rms noise in at least
2 consecutive channels.  The left-hand panels of Fig.~\ref{fig:fig1} show
the zeroth moment maps of the two systems; neither small galaxy
is well resolved, but the line emission from each galaxy is easily 
identifiable and separable in the \ion{H}{1} 
data cube.   In addition to the known 
galaxy pair, one additional galaxy was identified in the data 
cube of CGCG 132-062 at a heliocentric velocity of 2860 km s$^{-1}$; 
the coordinates correspond to the NED listing for MAP-NGP 0\_381\_0431360. 
We generated mock ``single dish'' \ion{H}{1} 
profiles for each galaxy to derive
the kinematic linewidths (Fig.~\ref{fig:fig1}).  The \ion{H}{1} linewidths
listed in Table \ref{tab:table} were measured at the 50\% of 
the peak points.

The 21cm lines of UGC~8919N and CGCG~132-062 are 
much wider than the collapsed major axis H$\alpha$ emission lines,
with $\sigma/\sigma_{\rm RADIO} = 0.67$ and 0.60, respectively.
The W$_{50}$ measurements are 102\% and 202\% 
wider, respectively, than the velocity widths (V$_{2.2}$)
of the resolved rotation curves.  
The \ion{H}{1} measurements shift the velocity widths upwards 
by $\sim$0.2~dex in Fig.~\ref{fig:re_sigma}, similar to the amount
predicted by the simple model of Fig.~\ref{fig:model}.  
The ``true'' (pre-burst) half-light radii 
may shift the galaxies into the spiral galaxy regime of 
Fig.~\ref{fig:re_sigma}.

\section{Tidally-Triggered Gas Infall and the Physics of Blue 
Compact Morphology}

Our study suggests that the central star formation
triggered by gas infall after a close galaxy-galaxy
pass is capable of causing discrepancies between 
\ion{H}{1} and emission-line kinematics.
Gas infall triggered by interactions \citep{MH96}
or minor mergers \citep{MH94} is analogous to secular 
evolution \citep{PN90}.  Thus, the process may explain the
``exponential bulges'' observed in some late-type spirals
\citep[e.g.,][]{AS94,C99}.
If hierarchical scenarios of galaxy formation are correct, 
triggered gas infall may occur frequently at intermediate
and high redshift.  The bulge-forming spirals --- especially 
the spirals with face-on inclinations or low surface brightness
disks --- would appear
similar to compact objects already observed at higher redshifts.

For local galaxies, existing studies shed little light on
the frequency of these T-F outliers among non-interacting
galaxies.  Most T-F studies would exclude the targets of this
study on the basis of morphology or because of poorly-sampled 
rotation curves.  The existing studies do show that 
centrally concentrated star formation is not a sufficient
condition for narrow emission lines. 
Several other galaxies in the \citet{B01} study
have centrally concentrated emission but are not
outliers to the T-F relation.  
One explanation for the anomalous properties of these 
outliers is that the systems may have particularly shallow
rotation curves.  Thus, the observed rotation
curve has a smaller total width when the new stars 
are centrally concentrated.
In any case, these local galaxies demonstrate
that centrally concentrated emission-line regions can, under
some circumstances, lead 
to anomalously narrow emission lines in disk galaxies.

In summary, we identify a set of four nearby galaxies with blue centers, small
half-light radii, and
anomalously narrow emission-line widths that are
possible counterparts to some of the compact blue galaxies observed
at intermediate redshift.  
\ion{H}{1} measurements, available for two of the galaxies,
show that the emission-line widths underestimate the kinematic widths
of the galaxies by $\gtrsim$50\%. Thus, by analogy, the 
emission-line widths of the compact objects at intermediate 
redshift may underrepresent their masses.
Because the local galaxies were selected from a sample
of pairs, the physical mechanism responsible for their
compact star formation is largely understood: a recent close galaxy-galaxy
interaction drove gas from the disk into the center of each galaxy,
triggering central star formation.  
This process, and the similar processes of minor
mergers and secular evolution, may facilitate bulge formation.
Thus, our study supports the hypothesis that at least some of
the luminous compact blue galaxies with narrow emission lines
observed at higher redshift are actually bulges forming in larger
systems.

\acknowledgements We thank John Salzer and Janice Lee for obtaining a
preliminary \ion{H}{1} 
spectrum of one of the targets for us in advance of the VLA
observations.  We thank Margaret Geller and David Crampton for useful
comments and suggestions.

\begin{deluxetable}{lcccccccc}
\tablenum{1}
\tablewidth{0pc}
\tablecolumns{9}
\tablecaption{Properties of the Local Sample}
\tablehead{
               &                       &                      & \colhead{R$_{\rm e}$} & \colhead{V$_{2.2}$ } & \colhead{V$^{\rm c}_{2.2}$} & \colhead{$\sigma$ } & \colhead{W$_{50}$ } & \colhead{W$_{50}^{\rm c}$} \\
\colhead{Name} & \colhead{M$_{\rm B}$} & \colhead{M$_{\rm R}$} & \colhead{(kpc)}    & \colhead{(km s$^{-1}$)} & \colhead{(km s$^{-1}$)} & \colhead{(km s$^{-1}$)} & \colhead{(km s$^{-1}$)} & \colhead{(km s$^{-1}$)} }
\startdata
UGC 8919N    & -18.3 & -19.4 & 3.52 & 50 & 64 & 29 & 101 & 112 \\
CGCG 132-062 & -18.0 & -19.0 & 1.87 & 40 & 42 & 31 & 121 & 127 \\
NGC 2719A    & -18.5 & -18.9 & 2.44 & 45 & 65 & 36 & ---  & ---\\
UGC 7085W    & -19.3 & -20.0 & 4.08 & 61 & 67 & 38 & ---  & ---\\
\enddata
\label{tab:table}
\end{deluxetable}

\end{document}